\documentclass[final,numberedheadings]{aipproc}

\layoutstyle{8x11double}

\newcommand{\apj}{Astrophysical Journal, }%
\newcommand{\apjl}{Astrophysical Journal Letters, }%
\newcommand{\apjs}{Astrophysical Journal Supplement, }%
\newcommand{\apss}{Astrophysics \& Space Science, }%
\newcommand{\mnras}{Monthly Notices of the Royal Astronomical Society, }%
\newcommand{\aap}{Astronomy \& Astrophysics, }%
\newcommand{\solphys}{Solar Physics, }%

\begin{document}

\title{A Stellar Model-fitting Pipeline for Solar-like Oscillations}

\author{T.~S. Metcalfe}{
address={High Altitude Observatory and Scientific Computing Division, 
NCAR, Boulder, Colorado, USA}}

\author{O.~L. Creevey}{
address={Instituto de Astrof{\'\i}sica de Canarias, 
La Laguna, Tenerife, Spain}}

\author{J.~Christensen-Dalsgaard}{
address={Department of Physics and Astronomy, Aarhus University,
Aarhus, Denmark}}

\begin{abstract}
Over the past two decades, helioseismology has revolutionized our 
understanding of the interior structure and dynamics of the Sun. 
Asteroseismology will soon place this knowledge into a broader context by 
providing structural data for hundreds of Sun-like stars. Solar-like 
oscillations have already been detected from the ground in several stars, 
and NASA's Kepler mission is poised to unleash a flood of stellar 
pulsation data. Deriving reliable asteroseismic information from these 
observations demands a significant improvement in our analysis methods. We 
report the initial results of our efforts to develop an objective stellar 
model-fitting pipeline for asteroseismic data. The cornerstone of our 
automated approach is an optimization method using a parallel genetic 
algorithm. We describe the details of the pipeline and we present the 
initial application to Sun-as-a-star data, yielding an optimal model that 
accurately reproduces the known solar properties.
\end{abstract}

\keywords{methods: numerical---stars: interiors---stars: oscillations}
\classification{97.10.Sj}

\maketitle


\section{Theorist's Perspective}

In 2004, during the SOHO-GONG meeting at Yale University, Art Cox gave a 
talk summarizing the attempts to identify g-mode oscillations in the Sun. 
He put up a slide showing an image of the GOLF instrument on the SOHO 
satellite and said, ``From a theorist's perspective, the light goes in 
here [pointing to the front end] and the answers come out here [pointing 
to the back end]''. We all know that there are actually a few more steps 
involved when analyzing and interpreting real data, but this paper 
describes a computational method that attempts to bring the modeling of 
solar-like oscillations one step closer to Art's idealized picture. 

In the past, ground-based data on solar-like oscillations in other stars 
have emerged slowly enough that we could try to model one star at a time. 
Beginning in October 2009, the Kepler mission promises to yield 
asteroseismic data for hundreds of stars every few months, so a hands-on 
approach will be a luxury we can no longer afford.


\section{Computational Method}

The basic idea behind our model-fitting pipeline is {\it frequencies in, 
stellar properties out}. The pipeline takes as input the observed 
oscillation frequencies and other constraints from non-seismic data. We 
use the Aarhus stellar evolution code \citep[ASTEC;][]{jcd08a} and the 
adiabatic pulsation code \citep[ADIPLS;][]{jcd08b} coupled with a parallel 
genetic algorithm (GA) to identify a global match between the models and 
the observations. We use the result of the global search as the starting 
point for a local analysis, which employs a modified Levenberg-Marquardt 
(LM) algorithm with Singular Value Decomposition (SVD) to determine the 
final parameter values and uncertainties, and to probe the information 
content of the observational constraints. The output of the pipeline 
includes the mass, initial composition, mixing length and stellar age, as 
well as other properties of the optimal model such as the temperature, 
luminosity and radius.

\subsection{Global Search \& Local Analysis}

Since we are interested in developing a general-purpose modeling tool for 
solar-like oscillations, we need to select a global method for optimizing 
the match between our model output and the available observations of any 
given star. Using only observations and the constitutive physics of the 
model to restrict the range of possible values for each parameter, a 
genetic algorithm \citep[GA;][]{cha95,mc03} can provide a relatively 
efficient means of searching globally for the optimal model. Although it 
is more difficult for a GA to find {\it precise} values for the optimal 
set of parameters efficiently, it is well suited to search for the {\it 
region} of parameter space that contains the global minimum. In this 
sense, the GA is an objective means of obtaining a good first guess for a 
more traditional local analysis method, which can narrow in on the precise 
values and uncertainties of the optimal model parameters.

Our implementation of the GA optimizes four adjustable model parameters; 
these are the stellar mass ($M_\star$) from 0.75 to 1.75 $M_\odot$, the 
metallicity ($Z$) from 0.002 to 0.05 (equally spaced in $\log Z$), the 
initial helium mass fraction ($Y_0$) from 0.22 to 0.32, and the mixing 
length parameter ($\alpha$) from 1 to 3. The stellar age ($\tau$) is 
optimized internally during each model evaluation by matching the observed 
mean separation $\left<\Delta\nu_0\right>$ between radial-mode frequencies 
(see section \ref{sec2.2}).

The GA uses two-digit decimal encoding, so there are 100 possible values 
for each parameter within the ranges specified above. Each run of the GA 
evolves a population of 128 models through 200 generations to find the 
optimal set of parameters, and we execute 4 independent runs with 
different random initialization to ensure that the best model identified 
is truly the global solution. This method requires about $10^5$ model 
evaluations, compared to $10^8$ models for a complete grid at the same 
sampling density, making the GA nearly 1000 times more efficient than a 
complete grid (currently 1 week of computing time, compared to many years 
for a grid). Of course, a grid could in principle be applied to hundreds 
of observational data sets without calculating additional models---but the 
GA approach also gives us the flexibility to improve the physical 
ingredients in the future, while the physics of a grid would be fixed.

Once the GA brings us close enough to the global solution, we can switch 
to a local optimization method. We implement a modified 
Levenberg-Marquardt (LM) algorithm that uses Singular Value Decomposition 
(SVD) to filter the least important information from the observables, some 
of which may be dominated by noise. 

We have three main motivations for implementing a local optimization 
method at the end of the global search. First, the GA has a limited 
resolution for each parameter, and the values that match the observations 
best are most likely between the fixed sample points. The resolution of 
the local analysis is limited only by the precision of the stellar 
evolution and pulsation codes, so we use it to adjust the models below the 
resolution of the GA search. Second, we need to quantify the final 
parameter uncertainties and correlations (not provided by the GA), and we 
want to probe the information content of the observables to determine 
which future observations can potentially help the most. Third, the local 
analysis can explore the effects of using different physical descriptions 
of the stellar interior \citep{cre08,cre09}. When the changes to the 
underlying physics are relatively subtle, we can assume that the global 
search by the GA using one set of assumptions will also provide a good 
starting point for a local analysis under slightly perturbed conditions.

\subsection{Fitting for Stellar Age\label{sec2.2}}

During the optimization process, each model evaluation involves the 
computation of a stellar evolution track from the zero-age main sequence 
through a mass-dependent number of internal time steps, terminating prior 
to the beginning of the red-giant stage. Rather than calculate the 
pulsation frequencies for each of the 200--300 models along the track, we 
exploit the fact that the average frequency spacing of consecutive radial 
overtones $\left<\Delta\nu_0\right>$ in most cases is a monotonically 
decreasing function of age \citep{jcd93}. Once the evolution track is 
complete, we start with a pulsation analysis of the model at the middle 
time step and then use a binary decision tree---comparing the observed and 
calculated values of $\left<\Delta\nu_0\right>$---to select older or 
younger models along the track. In practice, this allows us to interpolate 
the age between the two nearest time steps by running the pulsation code 
on just 8 models from each evolution track.

\subsection{Correcting for Surface Effects}

\begin{figure}
\includegraphics[width=0.48\textwidth]{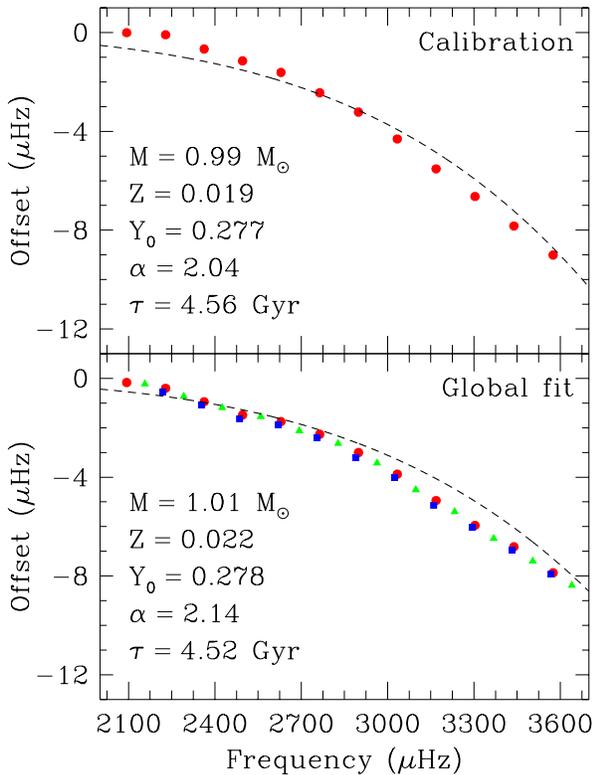}
\caption{{\em Top:} Surface effects lead to systematic offsets between the 
radial modes (circles) from BiSON data and our fit to the Model~S 
frequencies. {\em Bottom:} When we incorporate the resulting empirical 
correction into the models, applying it also to the dipole (triangles) and 
quadrupole (squares) modes, the global fit to the BiSON data reproduces 
the known solar properties within reasonable tolerances.\label{fig1}} 
\end{figure}

The biggest challenge to comparing the oscillation frequencies from 
theoretical models with those actually observed in solar-type stars are 
the systematic errors due to {\it surface effects}. The mixing length 
parameterization of convection that is used in most stellar models is 
insufficient to describe the near-surface layers, and this leads to a 
systematic difference of several $\mu$Hz (up to about 0.3\% for a solar 
model) between the observed and calculated frequencies (see 
Figure~\ref{fig1}). The offset is nearly independent of the spherical 
degree ($l$) of the mode and grows larger towards the acoustic cutoff 
frequency. The 3D simulations of convection that might in principle reduce 
this discrepancy for individual stars are far too computationally 
expensive for the model-fitting approach that we are developing. Instead, 
we adopt the method for empirical correction of surface effects described 
by \citeauthor{kbc08}\,\cite{kbc08}, which uses the discrepancies between 
Model S and GOLF data for the Sun \citep{laz97} to calibrate the empirical 
surface correction.

Following \citeauthor{kbc08}\,\cite{kbc08}, we fit a power law to the 
differences between the frequencies of the radial modes of our fit to 
Model S and the corresponding frequencies from BiSON data \cite{cha99} to 
characterize the surface effects. We found a power law exponent $b=4.82$, 
slightly lower than the value ($b=4.90$) derived by \citeauthor{kbc08} 
using data from the GOLF experiment. With this exponent fixed, the recipe 
of \citeauthor{kbc08} describes how to predict the surface effect for any 
other set of calculated oscillation data, allowing us to apply this 
empirical correction to each of our models before comparing them to the 
observations. If our strategy of making this empirical correction to each 
of our models is to succeed, it must not only work {\it well} for models 
in a certain region of the search space---it must work {\it best} for a 
model that simultaneously matches all of the independent observational 
constraints within their uncertainties.


\section{Initial Results}

Ultimately, our model-fitting pipeline can only be judged a success if it 
leads to accurate estimates of the stellar properties for the star that we 
know best: the Sun. There are many ingredients in our models that could in 
principle be insufficient descriptions of the actual conditions inside 
real stars---deficiencies that could easily lead to systematic errors in 
our determinations of the optimal model parameters for a given set of 
oscillation data. For example, we initially tried to use models that 
employed the simpler EFF equation of state \citep{eff73} for computational 
expediency, but this led to estimates of the stellar mass about 10\% too 
high for the Sun, and unacceptably large systematic errors on many of the 
other stellar properties. Even attempting to ignore the effects of helium 
settling proved to be too coarse an approximation, leading to 5\% errors 
on the mass. The only potential ingredient that we omitted without serious 
consequences was heavy element settling. This is not to say that simpler 
stellar models cannot be used in the analysis of asteroseismic data, but 
rather that some of the more sophisticated ingredients are required to 
obtain accurate results from a {\it global} search of the parameter space.

After demonstrating the effectiveness of the method by fitting our models 
to synthetic data and calibrating the empirical surface correction using 
the differences between our fit to Model S and the BiSON data 
\citep[see][]{mcc09}, we applied our model-fitting pipeline to solar data 
from the BiSON and GOLF experiments. The oscillation frequencies from 
these two sources are identical to each other within the observational 
uncertainties, but their noise properties are slightly 
different---allowing us to quantify any systematic errors that might arise 
from subtle effects in the data acquisition and analysis methods.

We assume that typical asteroseismic data from the Kepler mission will 
include twelve frequencies for each of the radial ($l=0$), dipole ($l=1$), 
and quadrupole ($l=2$) modes, with consecutive radial orders in the range 
$n=14{-}25$. Thus, we allowed the GA to fit a total of 36 oscillation 
frequencies. We assigned statistical uncertainties to each frequency by 
scaling up the errors on the corresponding modes by a factor of 10, which 
is roughly what we expect from Kepler data ($\sigma_\nu\sim0.1~\mu$Hz). We 
complemented this synthetic asteroseismic information with artificial data 
on the effective temperature and luminosity, with errors comparable to 
what is expected for stars in the Kepler Input Catalog \citep[][$T_{\rm 
eff}=5777\pm100$~K, $L_\star/L_\odot=1.00\pm0.1$]{lat05}. The two sets of 
input data differed only in the absolute values of the oscillation 
frequencies (yielding distinct values of $\left<\Delta\nu_0\right>$ for 
fitting the age), and in the statistical uncertainties assigned to each 
mode (leading to subtle differences in the weighting of the fit).

Both data sets led to identical values of the mass and metallicity from 
the global search, with slight variations in the values of the other 
parameters. These minor differences largely disappear after the local 
analysis. Note that because we multiplied the true observational errors by 
a factor of 10 for the fitting, the resulting values of $\chi^2_{\rm R}$ 
are $\sim$0.1. Although the fits used a limited range of frequencies and 
did not include $l=3$ modes, the optimal models also match the modes with 
lower frequencies and higher degree (see the BiSON fit in 
Figure~\ref{fig2}) and reproduce the known solar properties within 
reasonable tolerances.

\begin{figure}
\includegraphics[angle=270,width=0.48\textwidth]{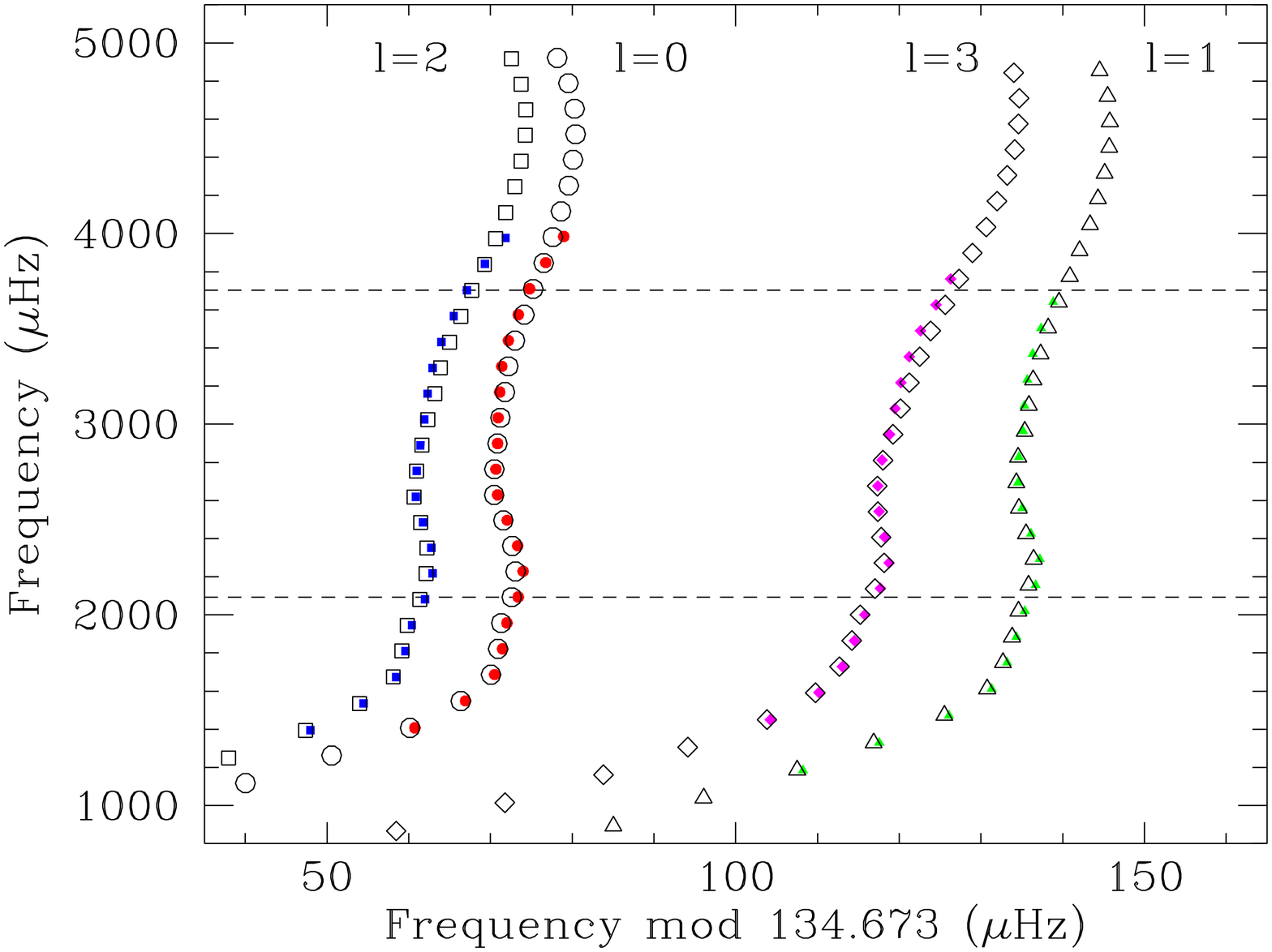}
\end{figure}
\begin{figure}
\includegraphics[angle=270,width=0.48\textwidth]{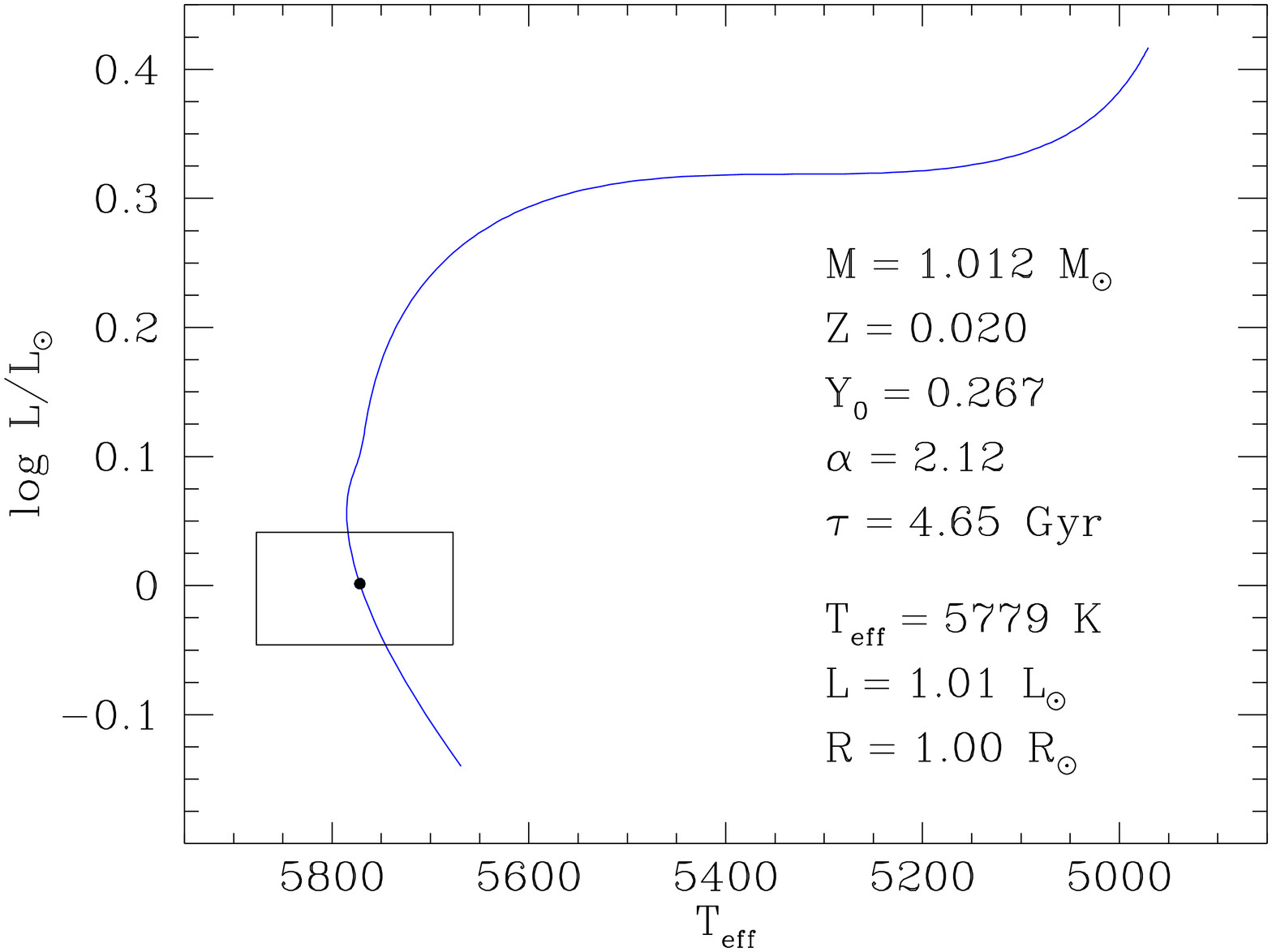}
\caption{{\em Top:} An echelle diagram for the BiSON data (solid points), 
where we divide the oscillation spectrum into segments of length 
$\left<\Delta\nu_0\right>$ and plot them against the oscillation 
frequency, along with the optimal model from our asteroseismic modeling 
pipeline (open points). Note that the pipeline only used the $l=0{-}2$ 
frequencies between the dashed lines for the fit, but the resulting 
optimal model also matches the $l$=3 modes and frequencies outside of the 
fitting range. {\em Bottom:} The evolution track (line) for the optimal 
model (solid point), which coincides with the middle of the observational 
error box (rectangle) defined by the temperature and luminosity 
constraints.\label{fig2}}
\end{figure}

\section{Future Work}

With the successful validation of our model-fitting pipeline using solar 
data, we now need to ensure that our adopted treatment of surface effects 
yields reasonable optimal models when applied to other stellar data. The 
obvious next step is to use archival ground-based data on several well 
studied solar-type stars to validate the pipeline for various stellar 
masses (e.g. $\alpha$ Cen A \& B, near $\sim$1.1 and $\sim$0.9 $M_\odot$ 
respectively) and for different evolutionary stages (e.g. the ``future 
Sun'' $\beta$~Hyi, at $\sim$7~Gyr). Although 
\citeauthor{kbc08}\,\cite{kbc08} demonstrated their method by applying it 
to models of these three stars in addition to the Sun, our experience 
using it with solar data suggests that a {\it global} exploration of the 
models may present additional challenges.

Since our stellar models and the empirical correction for surface effects 
have both been calibrated using a main-sequence star at 1.0~$M_\odot$, the 
$\alpha$~Cen system will help validate the models with interior physical 
conditions that differ slightly from those of the Sun. Its proximity and 
multiple nature make it an excellent second test of our pipeline, because 
it has very well determined properties including stellar radii from 
interferometry \citep{ker03}. There are also strong constraints on the 
component metallicities and effective temperatures \citep{plk08}, while 
the initial composition and age of the two stars are presumably identical. 
In the next phase of this project, we plan to use the published 
oscillation frequencies of $\alpha$~Cen~A \citep{bc02,bed04,baz07} and 
$\alpha$~Cen~B \citep{cb03,kje05} with the additional constraints from 
interferometry, spectroscopy, and the binary nature of the system to 
further validate our pipeline and the empirical correction for surface 
effects. \citeauthor{kbc08}\,\cite{kbc08} successfully applied their 
recipe to a set of stellar models that broadly resemble the components of 
the $\alpha$~Cen system, so we have good reason to believe that our 
implementation will also succeed---but this remains to be demonstrated.

The G2 subgiant $\beta$~Hyi has long been studied as a ``future Sun'',
with an age near 7~Gyr. It has been characterized almost as extensively as
the $\alpha$~Cen system, including recent interferometric measurements of
its diameter \citep{nor07} and dual-site asteroseismic observations
that determined its mean density with an accuracy of 0.6\% \citep{bed07}.
These data included the detection of several $l=1$ modes that deviate
from the asymptotic frequency spacing, suggesting that they are ``mixed
modes'' behaving like g-modes in the core and p-modes in the envelope.
This is expected for evolved stars like $\beta$~Hyi because as they expand
and cool the p-mode frequencies decrease, while the g-mode frequencies
increase as the star becomes more centrally condensed. This leads to a
range of frequencies where these modes can overlap and exchange their
character, manifested as so-called {\it avoided crossings}. This behavior
changes very quickly with stellar age, and propagates from one mode to the
next as a star continues to evolve. Consequently, the particular mode
affected yields a very strong constraint on the age of the star
\citep[see][]{jcd04}. In subsequent work, we plan to use the published
oscillation data for $\beta$~Hyi \citep{bed07} along with the constraints
from interferometry and spectroscopy to validate our pipeline and the
empirical treatment of surface effects for stars that are significantly
more evolved than the Sun. This will require an automated method to
recognize mixed modes in the data set and to incorporate them into the
optimization of stellar age along each track.

Once we have validated the model-fitting pipeline with additional stars
that sample a range of masses and evolutionary stages, we can begin to
consider additional observables and parameters that are not constrained by
currently available data sets. High-quality asteroseismic data are soon
expected from the Kepler mission, spanning sufficiently long periods of
time that the effects of rotation \citep{gs03,bal06,bal08} and magnetic
activity cycles \citep{cha07,met07} should be detectable. The Kepler
mission is designed to discover Earth-sized habitable planets, and our
model-fitting pipeline will be able to characterize the planet-hosting
stars with asteroseismology. This is essential to convert precise transit
photometry into an absolute radius for the planetary body. In addition,
accurate rotation rates and ages will provide clues about the formation
and evolution of the planet-hosting systems. The determination of accurate
stellar properties for a broad array of solar-type stars will give us a
new window on stellar structure and evolution, and will provide a broader
context for our understanding of the Sun and our own solar system. We hope
to facilitate this process by applying our stellar model-fitting pipeline
to the data that will soon emerge from the Kepler mission.


\begin{theacknowledgments}
The authors wish to thank Tim Brown and Margarida Cunha for helpful
discussions during the early phases of this project. This work was
supported in part by an NSF Astronomy \& Astrophysics Fellowship (to
T.S.M.) under award AST-0401441, by a Newkirk Graduate Fellowship (to
O.L.C.) at the High Altitude Observatory and by the European Helio- and
Asteroseismology Network (HELAS), a major international collaboration
funded by the European Commission's Sixth Framework Programme, by the
Danish Natural Science Research Council, and by NASA grant NNX09AE59G.
Computer time was provided by NSF MRI grants CNS-0421498, CNS-0420873,
CNS-0420985, the University of Colorado, and a grant from the IBM Shared
University Research (SUR) program. The National Center for Atmospheric
Research is a federally funded research and development center sponsored
by the U.S.~National Science Foundation.
\end{theacknowledgments}



\end{document}